\documentclass[a4paper,fleqn]{cas-dc}

\usepackage[numbers]{natbib}
\usepackage{todonotes}
\usepackage{hyperref}
\usepackage{multirow}

\begin{document}
\let\WriteBookmarks\relax
\def\floatpagepagefraction{1}
\def\textpagefraction{.001}

\shorttitle{Band Gap Reduction in Highly-Strained Silicon}

\shortauthors{N. Roisin et~al.}

\title [mode = title]{Band gap reduction in highly-strained silicon beams predicted by first-principles theory and validated using photoluminescence spectroscopy}

%
\author[1]{Nicolas Roisin}[type=editor,orcid=0000-0002-4251-1648]
\cormark[1]
\ead{nicolas.roisin@uclouvain.be}
\author[2]{Marie-Stéphane Colla}%
\author[1,3,4,5]{Romain Scaffidi}%
\author[2]{Thomas Pardoen}%
\author[1]{Denis Flandre}%
\author[1]{Jean-Pierre Raskin}%


\affiliation[1]{organization={ Institute of Information and Communication Technologies, Electronics and Applied Mathematics (ICTEAM), Université catholique de Louvain},
    addressline={Place du Levant 3}, 
    city={Louvain-la-Neuve},
    postcode={1348}, 
    country={Belgium}}
\affiliation[2]{organization={ Institute of Mechanics, Materials and Civil Engineering (IMMC), Université catholique de Louvain},
    addressline={Place du Levant 2}, 
    city={Louvain-la-Neuve},
    postcode={1348}, 
    country={Belgium}}
\affiliation[3]{organization={UHasselt - IMO-IMOMEC},
    addressline={Martelarenlaan 42}, 
    city={Hasselt},
    postcode={3500}, 
    country={Belgium}}
\affiliation[4]{organization={ imec - IMO-IMOMEC},
    addressline={Thorpark, Poort Genk 8310 \& 8320}, 
    city={Genk},
    postcode={3600}, 
    country={Belgium}}
\affiliation[5]{organization={EnergyVille - IMO-IMOMEC},
    addressline={Thorpark, Poort Genk 8310 \& 8320}, 
    city={Genk},
    postcode={3600}, 
    country={Belgium}}
\cortext[cor1]{Corresponding author}



\begin{abstract}
A theoretical study of the band gap reduction under tensile stress is performed and validated through experimental measurements. First-principles calculations based on density functional theory (DFT) are performed for uniaxial stress applied in the [001], [110] and [111] directions. The calculated band gap reductions are equal to 126, 240 and 100 meV at 2\% strain, respectively. Photoluminescence spectroscopy experiments are performed by deformation applied in the [110] direction. Microfabricated specimens have been deformed using an on-chip tensile technique up to  $\sim$ 1\% as confirmed by back-scattering Raman spectroscopy.
A fitting correction based on the band gap fluctuation model has been used to eliminate the specimen interference signal and retrieve reliable values. 
Very good agreement is observed between first-principles theory and experimental results with a band gap reduction of, respectively, 93 and 91 meV when the silicon beam is deformed by 0.95\% along the [110] direction.
\end{abstract}



\maketitle

\section{Introduction}

Silicon (Si) is a widely used material owing to its abundance, electronic performance and CMOS compatibility. The indirect band gap of Si around 1.12 eV at 300 K and its electronic properties make it suitable for visible and near-infrared photodetection\cite{Gao2017,Casalino2010,Liu2021} while the magnitude of its band gap limits the use for higher wavelength applications for which other materials such as indium gallium arsenide (InGaAs) or germanium (Ge) are preferred\cite{Fischer2022,Li2022}.

Elastic strain engineering is a proven technique to vary the electronic bandgap of semiconductors and to improve their charge carrier mobility at low additional costs. In the past years, theoretical works have shown the possibility to reduce the band gap and thus to extend the absorption range by deforming silicon crystals\cite{Roisin2021,Healy2014}. The enhancement of the optical and electrical performances of strained silicon makes it adapted to near-infrared sensing applications\cite{Katiyar2020}. 
The electronic properties of silicon under uniaxial or biaxial stress applied along the main crystal directions have been extensively studied using semi-empirical approaches based on fitting parameters such as the \textbf{k}$\cdot$\textbf{p} method\cite{Dresselhaus1955,Rideau2006}, empirical pseudopotential (EPM)\cite{Chelikowsky1976,Fischetti1996}  and tight-binding (TB) method\cite{Slater1954,Niquet2009}, or more recently using density-functional theory (DFT), allowing \textit{ab initio} computation without fitting parameters\cite{Bouhassoune2015,Wen2015,Shi2019,Janik2020}. Despite this potential and the existing first-principles studies, experimental works on the band gap reduction in deformed silicon remain scarce due to practical limitations of obtaining highly-deformed silicon samples and measuring the band structure of these specific samples. Furthermore, only a limited number of these works use their measurements to discuss theoretical studies.

Several experimental methods have been proposed in the literature to strain semiconductor-based devices without needing external actuation. Stressors\cite{Cavallo2012,Reiche2009,Wang2008, Wang2010} or epitaxial growth\cite{Cavallo2012,Reiche2009,Munguia2008a,Munguia2008b,Munguia2012} lead to a shift in the band gap and can help to significantly improve the mobility of the carriers in modern Si-based devices by applying uniaxial or biaxial strain respectively. The mobility gains are important at low strain levels and have been well-studied but the properties of highly-strained semiconductors remain partially explored. In\cite{Heremans2016}, transistor stacks were deformed in uniaxial tension and compression using bending while in\cite{Montmeat2016,sanchez2011direct} nanomembranes are biaxially strained using high pressure gas. Another strategy consists in using and amplifying the internal stress present in the semiconductor film\cite{gassenq20151}. All these techniques present limitation either in terms of level of applied strain and number of measurement points, or regarding the requirements for integration in a microelectronics fabrication process.

Only recently, it was shown how to reach high strains in different structures based on nanoribbons\cite{Passi2012,KumarBhaskar2013} and nanowires\cite{Passi2012,Urena2012,Zhang2016}. The nanoscale dimensions is key to reduce the number of defects in the strained structures and thus to allow reaching higher strain without failure. More specifically, on-chip tensile techniques allow the deformation of a large number of silicon beams without the need for external actuation\cite{Gravier2009,PARDOEN2016485,Bhaskar2011}. 

In this work, microfabricated silicon beams are deformed using an internal stress actuation principle. Raman spectroscopy is used to determine the elastic strain level while the band gap value is obtained by photoluminescence spectroscopy. As the vibrational modes of the crystal detected in the Raman spectrum depend on the structure of the semiconductor, the position of the Raman peaks can be directly linked to the intensity of the lattice deformation, i.e. elastic strain\cite{Ganesan1970,Roisin2023}. The luminescence spectrum collects the photons emitted by electron-holes recombination after excitation. The position of the spectrum and its shape are directly linked to the band structure and especially the band gap of the material\cite{Nguyen2017,Wagner1984,Yoo2015}.

Section \ref{sec:DFT} describes the first-principles simulations. The structural parameters of deformed silicon crystals are reminded, followed by the presentation of the computational methods. This section finished with a discussion about the theoretical results. Section \ref{sec:exp} describes the experimental work undertaken to validate the theoretical study. It starts with a description of the deformed silicon specimens and their fabrication process. Then, the  Raman spectroscopy method is explained, followed by the photoluminescence measurements. The results are finally discussed and compared at the end of Section \ref{sec:exp}.

\section{First-Principles simulations}\label{sec:DFT}

\subsection{Deformed silicon primitive cell}
\begin{figure}[h]
    \centering
    \includegraphics[width=\linewidth]{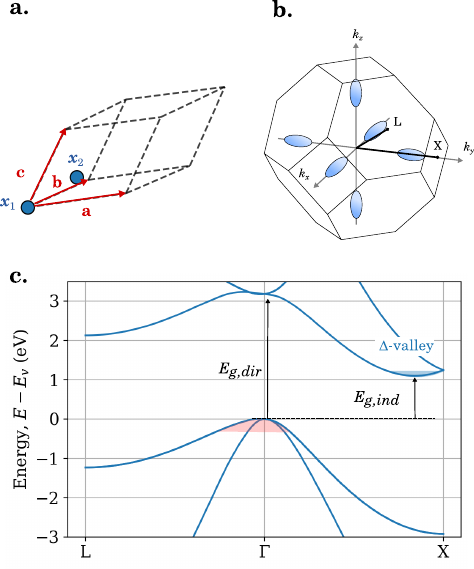}
    \caption{Representation of the two-atom primitive cell of silicon with (\textbf{a}) the lattices vectors and (\textbf{b}) the equivalent first Brillouin zone with the electron valleys. The path between the two symmetry points (L and X) is also represented (\textbf{c}) along with the band structure on this path with the minimum of the conduction band (blue area) and the maximum of the valence band (red area) highlighted. The direct band gap $E_{g,dir}$ and indirect band gap $E_{g,ind}$ are indicated.}
    \label{fig:Structure}
\end{figure}

The silicon unit cell is made of two inter-penetrating face-centered cubic (FCC) crystal structure with a two-atom primitive cell as displayed in Fig. \ref{fig:Structure}.a. The three Bravais lattice vectors are given by
\begin{equation}
    \mathbf{a}=\frac{a_0}{2}\left(\begin{array}{c}
         0  \\
         1 \\
         1
    \end{array}\right),~
        \mathbf{b}=\frac{a_0}{2}\left(\begin{array}{c}
         1  \\
         0 \\
         1
    \end{array}\right),~
        \mathbf{c}=\frac{a_0}{2}\left(\begin{array}{c}
         1  \\
         1 \\
         0
    \end{array}\right),
\end{equation}
where $a_0=0.543~\mathrm{nm}$ is the experimental lattice parameter\cite{Okada1984}.
The two atoms of the primitive cells are located at positions $\mathbf{x}_1$ and $\mathbf{x}_2$:
\begin{equation}
    \mathbf{x}_1=\left(\begin{array}{c}
         0  \\
         0 \\
         0
    \end{array}\right),\qquad
        \mathbf{x}_2=\left(\begin{array}{c}
         0.25  \\
         0.25 \\
         0.25\\
    \end{array}\right).\end{equation}
The reciprocal space presents a body-centered cubic (BCC) lattice. The first Brillouin zone is represented in Fig. \ref{fig:Structure}.b with the equipotential surface of the six electron valleys.

In a deformed crystal, the new set of lattice vectors $(\mathbf{a}', \mathbf{b}'$ and $\mathbf{c}')$ are given by
\begin{equation}
   (\mathbf{a}'~ \mathbf{b}'~ \mathbf{c}')= (\mathbf{I}+\textnormal{\textbf{E}})\cdot (\mathbf{a}~ \mathbf{b}~ \mathbf{c}),
\end{equation}
where $\mathbf{I}$ is the unit matrix and $\mathbf{E}$ is the strain tensor. In this work, uniaxial stress conditions applied along three crystal directions ([001], [110] and [111]) are studied theoretically. 
Under a tensile strain $\varepsilon_\perp$ applied along a crystal direction (or perpendicular to the crystal plane), the transverse dimensions reduce with Poisson ratio $\nu$ defined as

\begin{equation}
    \varepsilon_\parallel=-\nu~\varepsilon_\perp,
\end{equation}
where $\varepsilon_\parallel$ is the transverse strain.

The strain tensor and the important mechanical properties based on the components $C_{ij}$ of the compliance matrix are represented in Table \ref{tab:mech}.

\begin{table*}
\caption{Expressions of the strain tensor, Poisson ratio and Young's modulus corresponding to the loading applied in the [001], [110] and [111] directions where $C_{11}=165.77$~GPa, $C_{12}=63.93$~GPa, and $C_{44}=79.62$~GPa are the terms of the compliance matrice of silicon\cite{Madelung1991}. $\varepsilon_{\perp}$ stands for the tensile strain applied to the material while $\varepsilon_{\parallel}$ is the associated transverse strain.}
\label{tab:mech}
\begin{tabular}{cccc}\hline\hline
     \multicolumn{1}{c}{Strain} & Strain tensor & Poisson ratio & Young's Modulus  \\
               \multicolumn{1}{c}{direction}       & $\mathbf{E}$ & $\nu$ & Y  \\\hline
$[001]$& $\left(\begin{array}{ccc}
\varepsilon_{\parallel}&0&0\\
0&\varepsilon_{\parallel}&0\\
0&0&\varepsilon_{\perp}
\end{array}\right)$&$\frac{C_{12}}{C_{11}+C_{12}}$&$\frac{(C_{11}-C_{12})(C_{11}+2C_{12})}{(C_{11}+C_{12})}$\\ 
$[110]$&
$\frac{1}{2}\left(\begin{array}{ccc}(\varepsilon_{\perp}+\varepsilon_{\parallel})&(\varepsilon_{\perp}-\varepsilon_{\parallel})&0\\
(\varepsilon_{\perp}-\varepsilon_{\parallel})&(\varepsilon_{\perp}+\varepsilon_{\parallel})&0\\
0&0&2\varepsilon_{\perp}
\end{array}\right)$&$\frac{4C_{12}C_{44}}{2C_{11}C_{44}+(C_{11}+2C_{12})(C_{11}-C_{12})}$&$\frac{2}{\left(\frac{C_{11}}{C_{11}^2+C_{11}C_{12}-2C_{12}^2}+\frac{1}{2C_{44}}\right)}$\\
$[111]$& 
$\frac{1}{3}\left(\begin{array}{ccc}
(\varepsilon_{\perp}+2\varepsilon_{\parallel})&(\varepsilon_{\perp}-\varepsilon_{\parallel})&(\varepsilon_{\perp}-\varepsilon_{\parallel})\\
(\varepsilon_{\perp}-\varepsilon_{\parallel})&(\varepsilon_{\perp}+2\varepsilon_{\parallel})&(\varepsilon_{\perp}-\varepsilon_{\parallel})\\
(\varepsilon_{\perp}-\varepsilon_{\parallel})&(\varepsilon_{\perp}-\varepsilon_{\parallel})&(\varepsilon_{\perp}+2\varepsilon_{\parallel})\\
\end{array}\right) $&$\frac{C_{11}+2C_{12}-2C_{44}}{C_{11}+2C_{12}+2C_{44}}$&$\frac{3}{\left(\frac{C_{11}-C_{12}}{C_{11}^2+C_{11}C_{12}-2C_{12}^2}+\frac{1}{C_{44}}\right)}$\\\hline\hline
\end{tabular}

\end{table*}

\subsection{Computational methods}

The band structure is computed from first principles using density-functional theory (DFT) for relaxed and strained silicon, as implemented in ABINIT\cite{Gonze2020,Romero2020}. Norm-conserving pseudopotentials from the PSEUDO DOJO\cite{Vansetten2018} in the generalized-gradient approximation (GGA) from Perdew– Burke–Ernzerhof (PBE)\cite{Perdew1996}. A structural relaxation for the two atoms of the strained unit cell for each strain level is first performed. The electronic band structure is then determined using a Monkhorst–Pack k-point grid\cite{Monkhorst1976}. 
The convergence study performed on the relaxed silicon cell led to a band gap variation below 1 meV with a 18x18x18 grid and a plane-wave cutoff of 20 Ha. Many-body perturbation theory (MBPT), based on the Green's functions formalism (GW)\cite{Onida2002}, is used to correct the well-known band-gap underestimation in GGA\cite{Faleev2004}. In practice, quasi-particle (QP) energies and amplitudes are computed on a 10x10x10 k-point grid in the relaxed case to find the band gap correction.

\subsection{First-Principles results}
\begin{figure}[h]
    \centering
    \includegraphics[width=\linewidth]{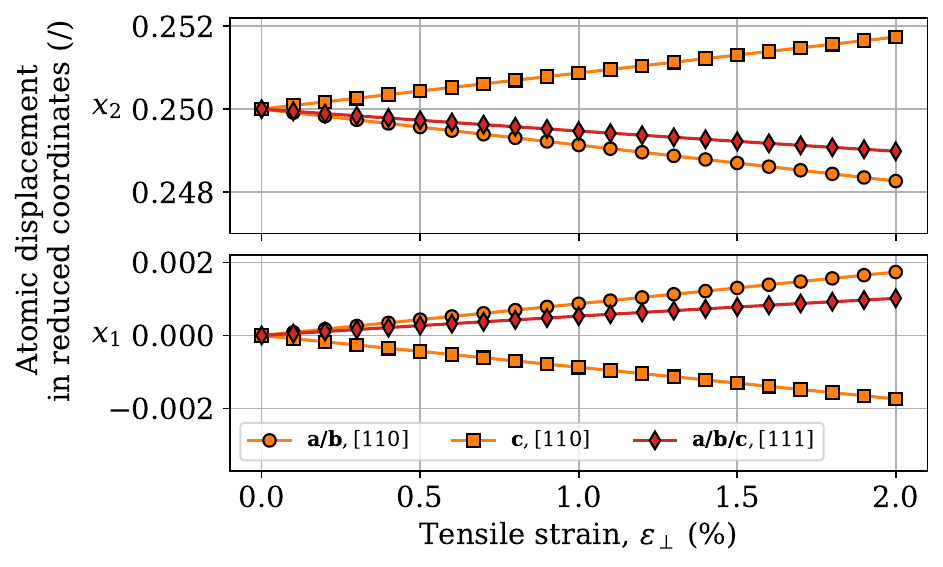}
    \caption{Relative positions of the atoms $x_1$ and $x_2$ inside the primitive cell under strain applied in the [110] (orange) and [111] (red) crystal direction. The indices \textbf{a},\textbf{b} and \textbf{c} refer to the lattice vectors.}
    \label{fig:displ}
\end{figure}

The results of the atomistic relaxation are shown in Fig. \ref{fig:displ}. Only the atomic displacement corresponding to a stress applied in the [110] and [111] directions are computed, as no internal relaxation happens in the [001] case due to the absence of shear distortion. In the first case, the structure displacement is directed along the lattice vector $\mathbf{c}$, causing an atomic displacement along this vector about  0.087 \% per unit of strain (in percent) (negative for $x_1$ and positive for $x_2$). The components for the other lattice vectors involve the same displacement but in opposite direction. In the [111] case, the displacement is 0.051 \% per unit of strain (in percent) and is similar along the three vectors (positive for $x_1$ and negative for $x_2$).

The band structure for relaxed silicon is represented in Fig. \ref{fig:Structure}.c. The electron valleys ($\Delta$), defined at the minima of the conduction bands, are six-fold degenerated along the $<$001$>$ directions. At these positions, the band gap around 0.57 eV determined using the PBE functional raises up to 1.12 eV with the GW correction. The 0.55 eV correction obtained in the relaxed case has been found to be hardly insensitive to the strain with a variation below 5 meV between the relaxed case and a tensile strain of 2\%. The slight change in the primitive cell due to the strain does not significantly impact the quasi-particle correction.

\begin{figure}[h]
    \centering
    \includegraphics[width=\linewidth]{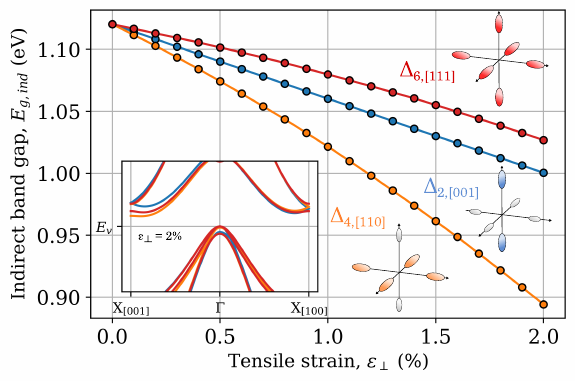}
    \caption{Indirect band gap reduction in silicon computed using density functional theory (DFT) for tensile strain applied in the [001] (blue), [110] (orange) and [111] (red) crystal direction. The degenerescence of the electron valleys is illustrated next to the curves with the lowest energy valleys highlighted. The inset represents the raw results of the band structure calculation for a 2\%-strain applied in the three crystal directions studied. The band structures are centered around the top of the valence band ($E_v$).}
    \label{fig:Bandgap}
\end{figure}

When a tensile strain is applied along the [110] or [001] direction, the broken symmetries of the crystal lift the degeneracy of the valleys into the $\Delta_2$-valleys along the [001] and [00$\bar{1}$] directions and the $\Delta_4$-valleys along [100], [010], [$\bar{1}$00] and [0$\bar{1}$0] directions. The variations of the valleys gaps are presented in Fig. \ref{fig:Bandgap}. In the [001] case, the $\Delta_2$ valleys present a reduced gap of 1.000 eV at 2\% while the gap of $\Delta_4$-valleys increases up to 1.224 eV. The opposite behavior is observed for [110] strain with lower gap energy down to 1.076 eV at 2\% for the $\Delta_2$ valleys 0.894 eV for the $\Delta_4$ valleys. On top of the reduction of the band gap, a split between the heavy-hole and light-hole valence bands up to 35 meV can be observed for a strain of 2\% oriented along the [110] direction.
The degeneracy is kept under the [111] tensile strain and the $\Delta_6$-valleys exhibit the smaller energy variations with respect to the strain (1.027 eV at 2\%). The important shear component coupled with a Poisson ratio about 0.36 make the [110] strain condition particularly suitable for band gap reduction compared to the [001] direction where no shear strain is present or to the [111] direction where limited variations of the crystal symmetries are induced. 
A slightly non-linear behavior is finally observed for the gap reductions, making the first-principles theory necessary to correctly  studythe effects at high strain levels.

\section{Experimental validation}\label{sec:exp}

\subsection{Fabrication of deformed silicon microbeams}

An on-chip tensile testing technique has been used to stretch silicon microbeams without external actuation\cite{Gravier2009,PARDOEN2016485}. The concept of this technique is to benefit from the high internal stress present in an \textit{actuator} beam made here of silicon nitride in order to pull on the \textit{specimen} microbeam, made of silicon in this study\cite{Passi2012,KumarBhaskar2013,Urena2012,Bhaskar2011,Escobedo-Cousin2011}.

\begin{figure}[h]
    \centering
    \includegraphics[width=\linewidth]{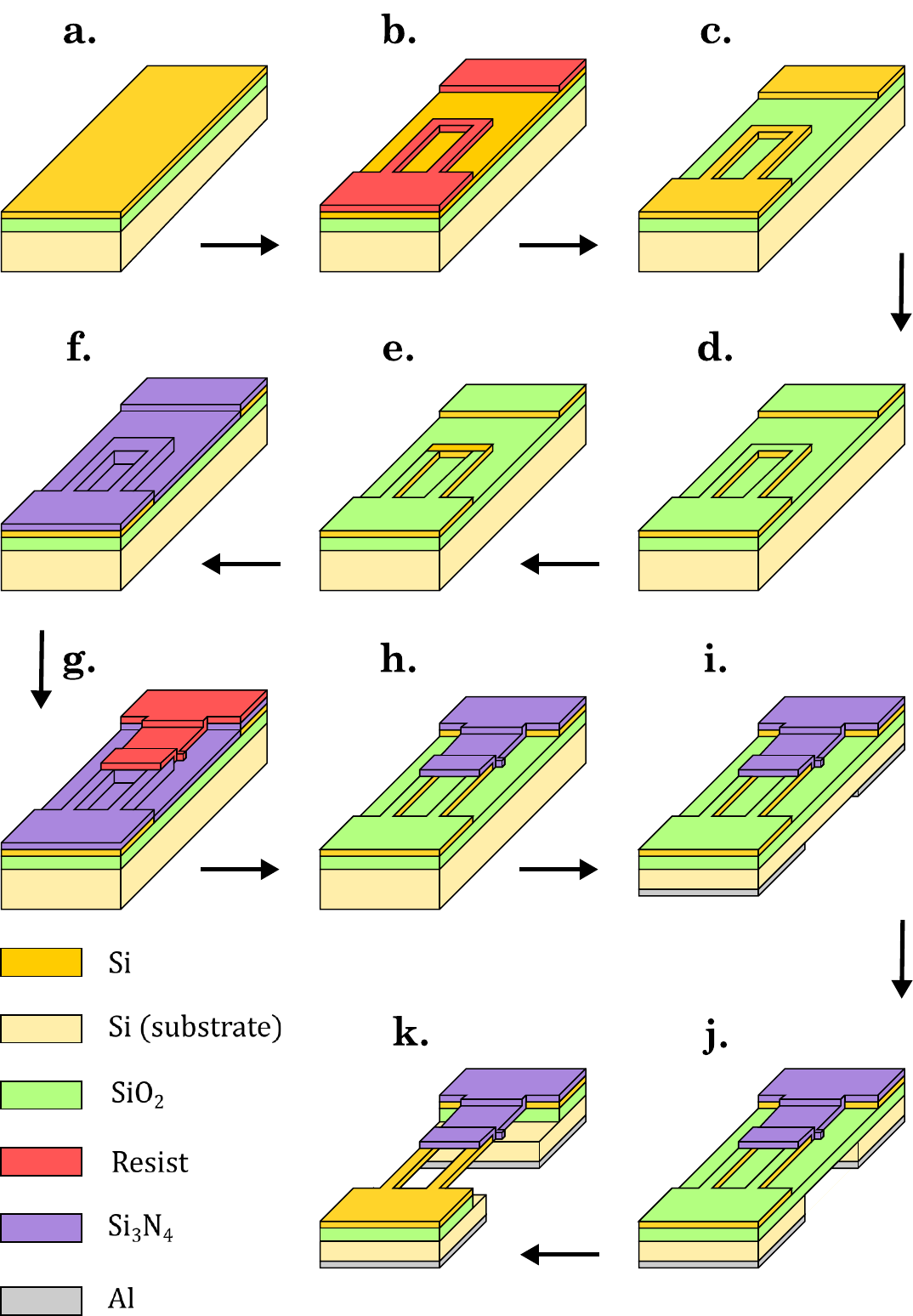}
    \caption{Overview of the fabrication process of strained silicon microbeams. The silicon beams are patterned first (a$\rightarrow$c), followed by the silicon nitride patterning (d$\rightarrow$h). The back etching and release of the structures are performed at the end of the process (i$\rightarrow$k).}
    \label{fig:Process}
\end{figure}

The design of the on-chip tensile testing is shown in Fig. \ref{fig:Process}. Microfabrication techniques as developed for microelectronics are used to design and build the test structures. The fabrication process starts with a p-type 700 µm-thick Silicon-on-Insulator (SOI) substrate with a 685 nm-thick top Si layer and a 1 µm-thick buried oxide (BOX) that will serve as a sacrificial layer. First, the top Si layer is patterned thanks to a photolithography step followed by a CHF$_3$/SF$_6$ plasma etching to shape to the microbeam specimens (Figs. \ref{fig:Process} a to c). Then, a 20 nm-thick dry thermal oxide layer is grown at 950°C on the patterned top-Si and locally etched away after photolithography using BHF solution in order to open windows at the overlapping area where the actuator is in contact with the top Si microbeams (Figs. \ref{fig:Process} d to e). Finally, the actuator layer made of low pressure chemical vapour deposited (LPCVD) silicon nitride film is deposited at 800°C on top of the partly oxidized top-Si and patterned using CHF$_3$/O$_2$ dry etching (Figs. \ref{fig:Process} f to h). The 20 nm-thick  oxide on the top Si serves as an etch stop layer during this processing step, thus avoiding the degradation of the Si beams. The fabrication process of the top-face of the substrate is completed as shown in Fig. \ref{fig:Process} h. For characterization purposes, as in this work, it is needed to open windows in the bulk Si substrate under the microbeams of interest using deep reactive ion etching following a previous study\cite{Colla2015}. 

The back-etching step begins with a thinning of the whole substrate by mechanical polishing to reduce the substrate thickness down to 250 µm. Then, a 300 nm-thick Al layer is evaporated and patterned using photolithography followed by H$_3$PO$_4$ etching. This Al layer acts as a hard mask during deep reactive ion etching (Fig. \ref{fig:Process} i). The front face of the wafer containing the microbeams is protected by photoresist. The Bosch process is used for the deep Si etching. The cyclic etching is performed by alternating 12 seconds-long SF$_6$ etching steps and 13 seconds-long C$_4$F$_8$ deposition steps. 500 cycles have been applied to fully open the windows in the substrate. The buried oxide sacrificial layer is acting as an etch-stop layer and ensures that the microbeams are not affected by the back-etching process (Fig. \ref{fig:Process} j).

\begin{figure}[h]
    \centering
    \includegraphics[width=\linewidth]{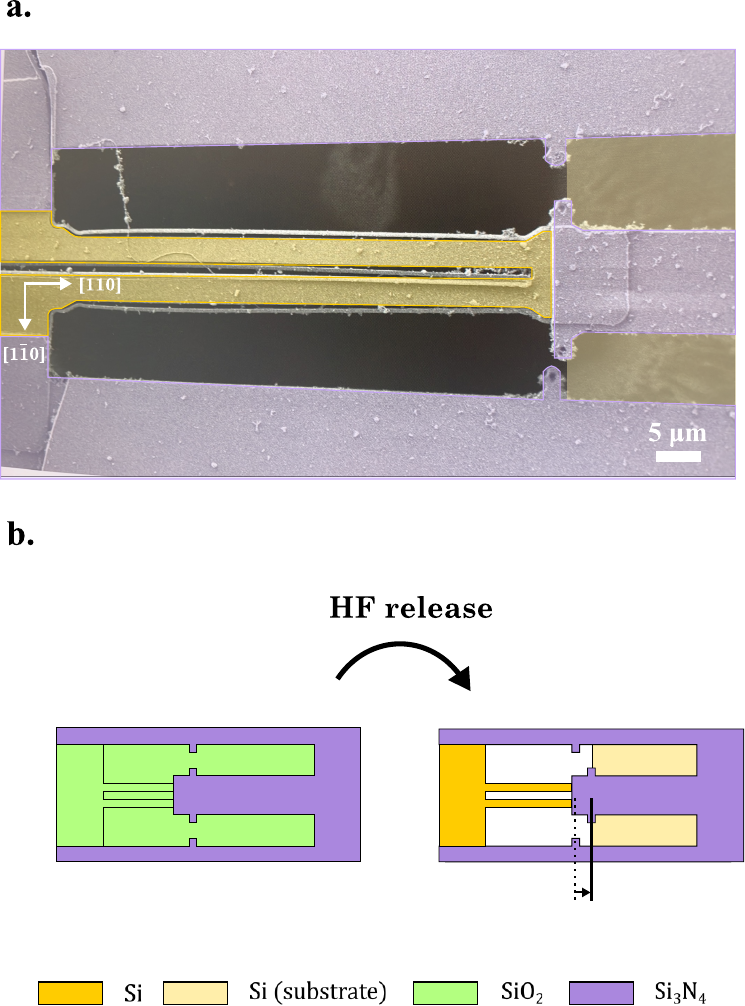}
    \caption{Scanning Electron Microscopy (SEM) micrograph of a deformed microbeams structure (\textbf{a.}). Stress is applied after the silicon oxide (SiO$_2$) etching with a hydrofluoric acid solution concentrated at 49\% (\textbf{b.}). The silicon beams are pulled under tension due to the internal stress in the silicon nitride (Si$_3$N$_4$) actuator.}
    \label{fig:SEM}
\end{figure}

The last step is the release of the entire structure using hydrofluoric acid (HF) concentrated at 49\% in order to remove the sacrificial buried oxide lying under the specimen and actuator beams and in order to let the force equilibrium to operate (Figs. \ref{fig:Process} k). After the release, the Si$_3$N$_4$ beam is pulling on the Si microbeam that is deformed under tension. The strain in each Si beam can be calculated from the relative displacement of the freestanding structure thanks to a series of fixed and moving cursors\cite{Gravier2009} or by Raman spectroscopy as in this work (see next section). Hundreds of such elementary structures with varying dimensions are produced on one die. The length of the actuator beams is varied from 100 to 1500 µm to access to different strain levels inside the Si microbeams that have a constant length of 50 µm. The width of the actuator beams is equal to 10 µm while the Si microbeams are 2 µm-wide.

\subsection{Strain determination by Raman spectroscopy}

\begin{figure}[h]
    \centering
    \includegraphics[width=\linewidth]{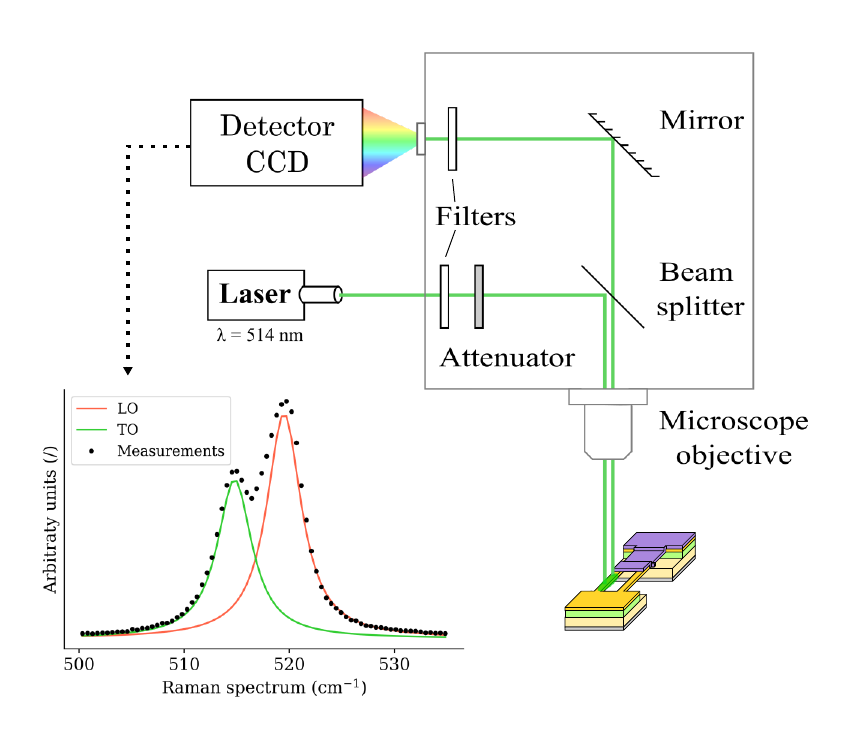}
    \caption{Raman spectroscopy set-up to determine the intensity of the strain field in the deformed silicon samples. A 514 nm laser is guided onto the sample and the scattered signal is collected through a beamsplitter and diffraction grating to be analyzed on a charge-coupled device (CCD). The spectrum is then analyzed by fitting Lorentzian peaks to extract the position of the LO and TO phonon modes that depend on the magnitude of the elastic strain. }
    \label{fig:Raman}
\end{figure}

Raman spectroscopy is a well-known non-invasive method to extract the magnitude of the elastic strain field in a semiconductor based on the vibration modes of the crystal\cite{DeWolf1996}.
In silicon, the degeneracy of the three optical phonon peaks (around 520.7 cm$^{-1}$) is lifted under deformation. The intensity of the strain field can then be directly determined from the peak positions if the proper relations between the strain and the shifts of the peaks are known.

The labRAM HR system from \textit{Horiba} with a 514 nm laser was used to perform backscattering Raman spectroscopy. The system, illustrated in Fig. \ref{fig:Raman}, is based on a confocal microscope with a filter and an attenuator on the incident light path while a spectrograph grating coupled with mirrors discriminates spatially the energy of the scattered light on a charge-coupled device (CCD). 

According to the Raman selection rule, only the LO mode should be excited with the laser polarized in the [110] direction across the (001) crystal plane\cite{Loudon1964}. However, the high numerical aperture (NA) of the microscope objective (MPLAPON100x from \textit{Olympus}) leads to the observation of TO mode thanks to the tilted components of the light beam hitting the material. 
Indeed, input light with an incident angle up to 70° can be collected with a numerical aperture of 0.95. The spectrum can then be fitted with Lorentzian functions to retrieve the positions of the Raman peaks\cite{Pelikan1994}.

\begin{figure}[h]
    \centering
    \includegraphics[width=\linewidth]{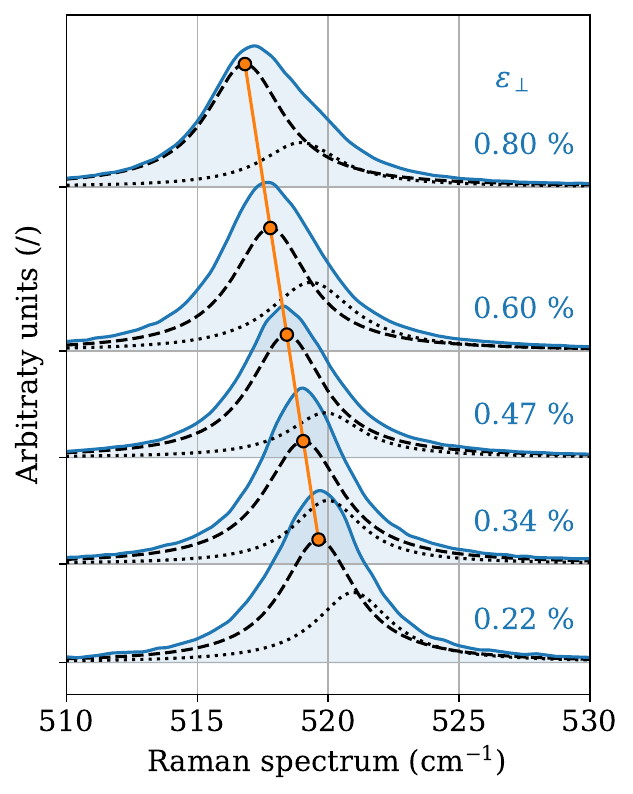}
    \caption{Raman spectrum of several deformed microbeams. The measurement data (continuous blue lines) are fitted using two Lorentzian functions representing the contribution of the TO mode (dashed lines) and LO mode (dotted lines). The spectra are normalized with regards to the intensity of the TO peaks while the vertical position of each spectrum is related to the strain of the sample. This strain level has been determined using the linear relationship (continuous orange line) between the peak position of the TO modes (oranges dots) and the strain.}
    \label{fig:Raman_spectrum}
\end{figure}

 Fig. \ref{fig:Raman_spectrum} shows the Raman spectra of five strained samples from 0.22\% to 0.80\%. The Lorentzian fit representing the two phonon modes (TO and LO) is highlighted while the vertical position is set according to the strain of the samples. The strain level has been computed using the position of the TO peaks with a strain-shift coefficient of -485 cm$^{-1}$\cite{Urena2013}.
About ten measurement points are taken along each silicon beam in order to determine the uncertainty of the measurements.

\subsection{Photoluminescence measurements}

\begin{figure}[h]
    \centering
    \includegraphics[width=\linewidth]{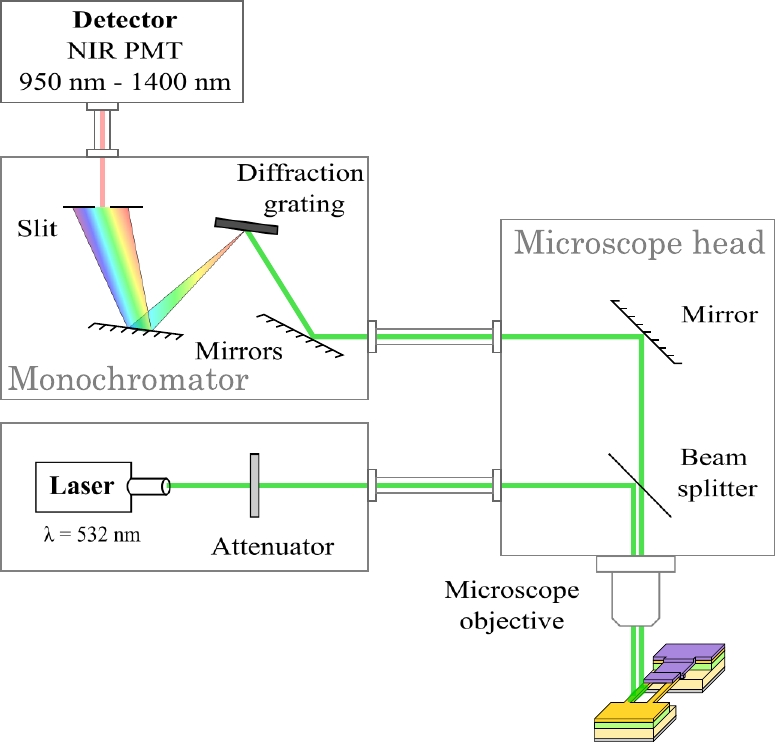}
    \caption{Diagram of the photoluminescence set-up used to determine the band gap of the deformed microbeams. A 532-nm laser has been used with confocal backscattering configuration. The signal emitted by the sample is then collected through the same objective as the input laser beam. A monochromator coupled to a NIR photodetector is used at the end to retrieve the luminescence spectrum.}
    \label{fig:PLsystem}
\end{figure}

The FluoMic and FluoTime 300 from \textit{Picoquant} are used to generate the luminescence spectrum of the strained samples as represented in Fig. \ref{fig:PLsystem}. A 532 nm pulsed laser going through an attenuator is focused on the sample with a MPLN50x objective from \textit{Olympus}. The spot size of about 20 $\mathrm{\mu}$m diameter leads to a nearly complete illumination of the deformed samples. The duration of the pulse is 25 ps with a 3 MHz repetition rate. The back-scattering system collects the emitted light through the same objective. A beam splitter separates the input laser line from the emitted spectrum, guided to a monochromator. The monochromatic output light is analyzed at the end by a NIR photodetector (900 nm - 1400 nm) with photomultiplier tube (PMT). The full spectrum is then obtained by tuning the monochromator between 900 and 1300 nm with step of 1 nm. The integration time for each point of the spectrum is set to 15 s. 


\begin{figure}[h]
    \centering
    \includegraphics[width=\linewidth]{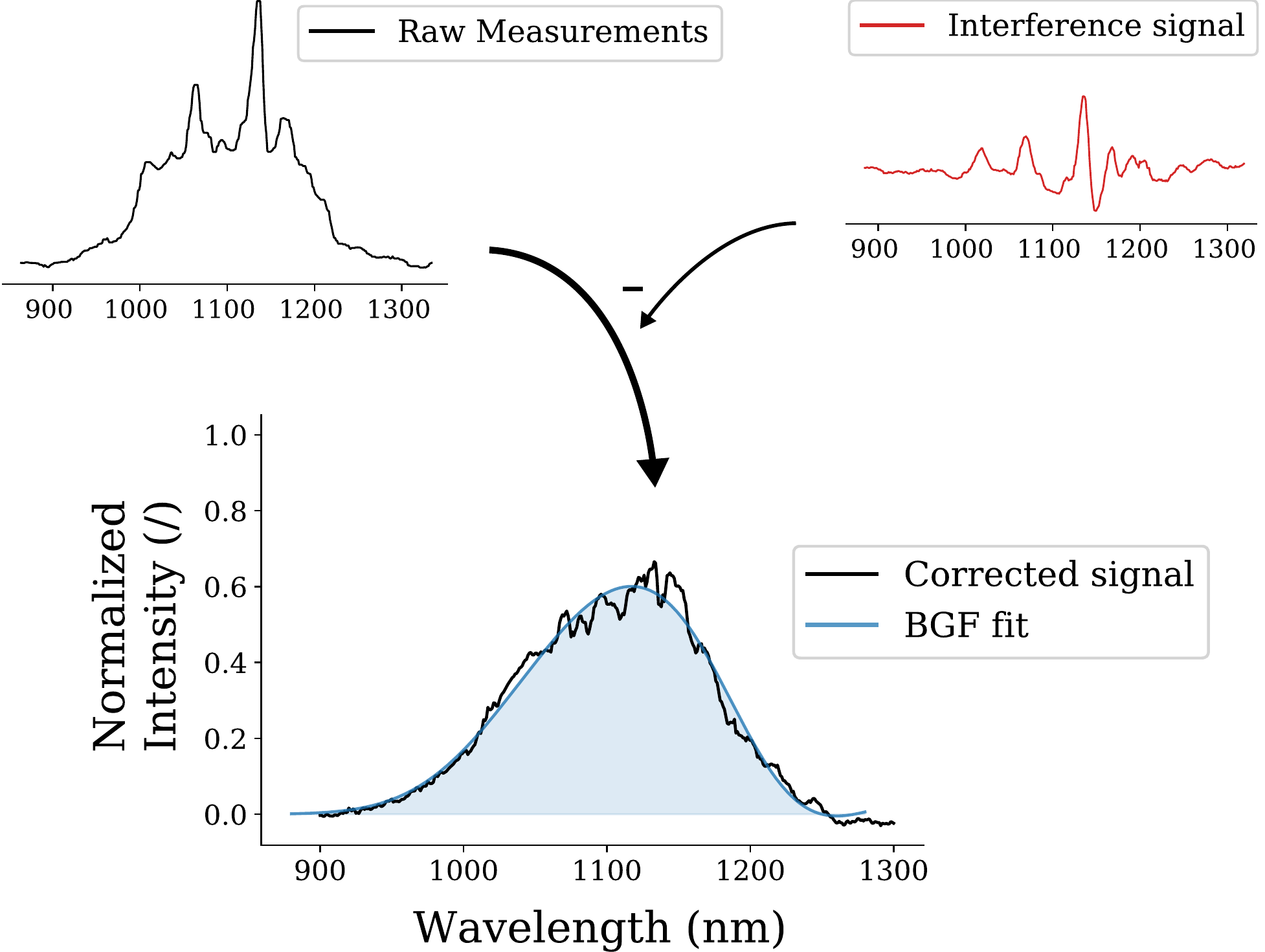}
    \caption{Data processing scheme to remove the interference signal (red) from the raw measurements (black) before the fit with the photoluminescence model (blue).}
    \label{fig:DataProcessing_bis}
\end{figure}

The luminescence spectrum $\Phi(E)$ is fitted and calibrated using the band gap fluctuation model\cite{Rau2004,Mattheis2007}:
\begin{equation}
    \Phi(E)\propto \mathrm{erfc}\left(\frac{\bar{E_g}-E}{\sqrt{2}\sigma_g}\right)E^2\mathrm{exp}\left(-\frac{E-\mu}{k_B T}\right),
\end{equation}
where $E$ is the photon energy, $\mu$ is the chemical potential, $k_BT$ is the thermal energy, $\sigma_g$ is the band gap fluctuation, $\bar{E_g}$ is the mean band gap.
A first fit is performed on the sample to separate the interference signal from the luminescence signal. A second fit is then applied to the signal for which the interferences have been subtracted to retrieve the final band gap value. The correction and fitting steps are represented in Fig. \ref{fig:DataProcessing_bis}. The back-etched configuration of the microfabricated samples avoids a background signal from the silicon substrate and thus improves the quality of the measurements.
Thirteen beams have been characterized with strain levels between 0.2\% and 1\%, leading to a band gap reduction between 22 meV and 91 meV.

\begin{figure}[h]
    \centering
    \includegraphics[width=\linewidth]{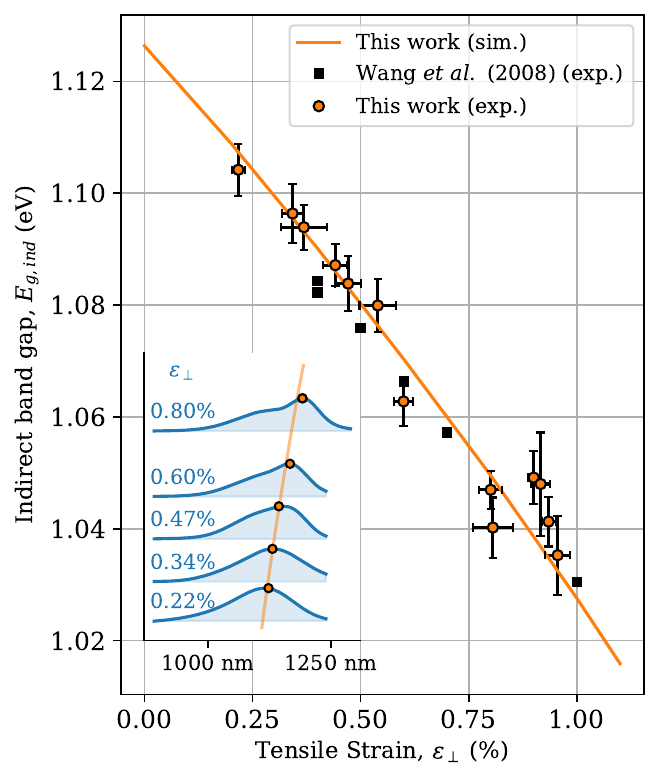}
    \caption{Results of the band gap determination from Raman and photoluminescence measurements (orange dots). The vertical and horizontal bars around the experimental results are the strain and band gap uncertainties coming from the Raman and the luminescence measurements, respectively. Data are compared with experimental results found in the literature (black square)\cite{Wang2008} and to the simulation results obtained by density-functional theory (orange line). The inset represents the PL spectrum fitted on experimental data after the post-processed step to remove the interference signal. The band gap position for different strain levels is indicated on the spectrum (orange dot) as well as the first-principles prediction (orange line).}
    \label{fig:PL_results}
\end{figure}

In order to estimate the uncertainty due to the data processing, a statistical approach has been chosen with corrections performed on each Si beam with the interference signal from the other ones. The results of the data processing are shown in Fig. \ref{fig:PL_results}. The coordinates of each point refer to the mean value for the band gap and strain level determined for each Si beam. The vertical bars represent the PL uncertainty of the band gap due to the interference corrections. For each deformed beam, the PL measurements are fitted after removing the interference signal from all beams. The size of the bars is equal to three times the standard deviation of the band gap values extracted after the interference corrections.
The Raman shift at eleven points along each beam is extracted to compute the uncertainty on the strain, represented by the horizontal bars. The width of the bars is equal to three times the standard deviation coming from the strain computed on each point of the test beams.

The inset of Fig. \ref{fig:PL_results} shows normalized luminescence spectra with the peak position highlighted for several deformation levels and compared with the theoretical results.

The first-principle results for the tensile strain applied along the [110] directions have been represented and show good agreement with the experimental results. The largest band gap reduction found experimentally is about 91 meV at 0.95 \% while a reduction of 93 meV has been predicted at this level of deformation. The results extend a previous experimental determination using PL spectroscopy\cite{Wang2008} where a band gap reduction of 100 meV has been found for a strain about 1\%. Previous study showed a band reduction of 69 meV for a strain of $\sim$ 0.6\% applied in the [100] crystal direction\cite{Wang2010}. Higher band gap reduction of 239 meV was obtained with a biaxial strain of 1.54\% applied on a (001)-silicon layer using Si$_x$Ge$_{1-x}$ stressor\cite{Munguia2008a,Munguia2008b}.

This work experimentally demonstrates the physical insight given by the first-principles simulations to characterize the band gap reduction in highly-strained silicon. 
On top of that, the results of the PL and Raman measurements confirm the use of strain engineering to tune the optical behavior of deformed materials. Further investigation will address different loading configurations involving shear stress, uniaxial and biaxial stress in other crystal directions or with compressive structures instead of tensile ones.

\section{Conclusion}

In this work, a high band gap reduction of 240 meV under 2\% strain was found by density-functional theory (DFT) for tensile stress applied in the [110] direction of the silicon cell. The first-principles simulations have been confirmed experimentally thanks to on-chip tensile method with back-etched substrate to minimize background interference during the measurements. 
Silicon microbeams deformed up to 1\% have been observed using a photoluminescence system. The final results give a maximum reduction of 91 meV at a strain level of 0.95\%, showing good agreement with the theoretical analysis. 

By combining the results of the first-principles study with the experiments, this work confirms the theoretical understanding of band structure tailoring under stress conditions, making silicon an interesting material for infrared applications.
The addition of the DFT simulation to the PL measurements for investigating the reduction of the band gap provides a deeper and more nuanced understanding of the behavior of highly-strained microbeams.
The findings of this work highlight the potential of strain engineering as a powerful tool for tailoring the properties of semiconductors at small scales and suggest that further research in this area could lead to new possibilities for optoelectronic devices and applications.

\section*{Acknowledgments}

Computational resources have been provided by the supercomputing facilities of the Université catholique de Louvain (CISM/UCL) and the Consortium des Équipements de Calcul Intensif en Fédération Wallonie Bruxelles (CÉCI) funded by the Fond de la Recherche Scientifique de Belgique (F.R.S.-FNRS) under convention 2.5020.11 and by the Walloon Region. M.-S. Colla acknowledges the financial support of National Fund for Scientific Research (FNRS), Belgium. N. Roisin acknowledges the help received for the first-principles simulations from G.-M. Rignanese and G. Brunin from Institute of Condensed Matter and Nanosciences (IMCN), Belgium. 
The authors acknowledge the help received from IMO-IMOMEC, the joint department of imec and UHasselt, to realize the photoluminescence measurements on the site of EnergyVille in Genk, Belgium, and especially the support provided by Dr. Guy Brammertz with the experimental setup and the EU ERC project (Uniting PV, H2020 research and innovation programme under grant agreement n°715027) of Prof. Bart Vermang that funded the equipment.

\section*{Competing Interests}
The authors have no relevant financial or non-financial interests to disclose.

\section*{Data Availability}
All data generated or analysed during this study are included in this published article.


\bibliographystyle{unsrtnat}

\bibliography{main}

\providecommand{\noopsort}[1]{}\providecommand{\singleletter}[1]{#1}%
\begin{thebibliography}{58}
\providecommand{\natexlab}[1]{#1}
\providecommand{\url}[1]{\texttt{#1}}
\expandafter\ifx\csname urlstyle\endcsname\relax
  \providecommand{\doi}[1]{doi: #1}\else
  \providecommand{\doi}{doi: \begingroup \urlstyle{rm}\Url}\fi

\bibitem[Gao et~al.(2017)Gao, Cansizoglu, Polat, Ghandiparsi, Kaya, Mamtaz,
  Mayet, Wang, Zhang, Yamada, Devine, Elrefaie, Wang, and Islam]{Gao2017}
Yang Gao, Hilal Cansizoglu, Kazim~G. Polat, Soroush Ghandiparsi, Ahmet Kaya,
  Hasina~H. Mamtaz, Ahmed~S. Mayet, Yinan Wang, Xinzhi Zhang, Toshishige
  Yamada, Ekaterina~Ponizovskaya Devine, Aly~F. Elrefaie, Shih~Yuan Wang, and
  M.~Saif Islam.
\newblock {Photon-trapping microstructures enable high-speed high-efficiency
  silicon photodiodes}.
\newblock \emph{Nature Photonics}, 11\penalty0 (5):\penalty0 301--308, 2017.
\newblock ISSN 17494893.
\newblock \doi{10.1038/nphoton.2017.37}.

\bibitem[Casalino et~al.(2010)Casalino, Coppola, Iodice, Rendina, and
  Sirleto]{Casalino2010}
Maurizio Casalino, Giuseppe Coppola, Mario Iodice, Ivo Rendina, and Luigi
  Sirleto.
\newblock {Near-Infrared Sub-Bandgap All-Silicon Photodetectors: State of the
  Art and Perspectives}.
\newblock \emph{Sensors}, 10\penalty0 (12):\penalty0 10571--10600, nov 2010.
\newblock ISSN 1424-8220.
\newblock \doi{10.3390/s101210571}.
\newblock URL \url{http://www.mdpi.com/1424-8220/10/12/10571}.

\bibitem[Liu et~al.(2021)Liu, Cristoloveanu, and Wan]{Liu2021}
Jian Liu, Sorin Cristoloveanu, and Jing Wan.
\newblock {A Review on the Recent Progress of Silicon‐on‐Insulator‐Based
  Photodetectors}.
\newblock \emph{physica status solidi (a)}, 218\penalty0 (14), jul 2021.
\newblock ISSN 1862-6300.
\newblock \doi{10.1002/pssa.202000751}.
\newblock URL
  \url{https://journals.aps.org/rmp/abstract/10.1103/RevModPhys.74.601}.

\bibitem[Fischer et~al.(2022)Fischer, Brehm, {De Seta}, Isella, Paul, Virgilio,
  and Capellini]{Fischer2022}
I.~A. Fischer, M.~Brehm, M.~{De Seta}, G.~Isella, D.~J. Paul, M.~Virgilio, and
  G.~Capellini.
\newblock {On-chip infrared photonics with Si-Ge-heterostructures: What is
  next?}
\newblock \emph{APL Photonics}, 7\penalty0 (5), may 2022.
\newblock ISSN 2378-0967.
\newblock \doi{10.1063/5.0078608}.
\newblock URL \url{https://doi.org/10.1063/5.0078608
  https://aip.scitation.org/doi/10.1063/5.0078608}.

\bibitem[Li et~al.(2022)Li, Zhang, Yue, Tang, Gao, Jiang, Du, Deng, Jia, Wang,
  and Chen]{Li2022}
Xuanzhang Li, Junyang Zhang, Chen Yue, Xiansheng Tang, Zhendong Gao, Yang
  Jiang, Chunhua Du, Zhen Deng, Haiqiang Jia, Wenxin Wang, and Hong Chen.
\newblock {High performance visible-SWIR flexible photodetector based on
  large-area InGaAs/InP PIN structure}.
\newblock \emph{Scientific Reports}, 12\penalty0 (1), may 2022.
\newblock ISSN 2045-2322.
\newblock \doi{10.1038/s41598-022-11946-7}.
\newblock URL \url{https://doi.org/10.1038/s41598-022-11946-7
  https://www.nature.com/articles/s41598-022-11946-7}.

\bibitem[Roisin et~al.(2021)Roisin, Brunin, Rignanese, Flandre, and
  Raskin]{Roisin2021}
Nicolas Roisin, Guillaume Brunin, Gian-Marco Rignanese, Denis Flandre, and
  Jean-Pierre Raskin.
\newblock {Indirect light absorption model for highly strained silicon infrared
  sensors}.
\newblock \emph{Journal of Applied Physics}, 130\penalty0 (5), aug 2021.
\newblock ISSN 0021-8979.
\newblock \doi{10.1063/5.0057350}.
\newblock URL \url{https://aip.scitation.org/doi/10.1063/5.0057350}.

\bibitem[Healy et~al.(2014)Healy, Mailis, Bulgakova, Sazio, Day, Sparks, Cheng,
  Badding, and Peacock]{Healy2014}
Noel Healy, Sakellaris Mailis, Nadezhda~M. Bulgakova, Pier~J.A. Sazio, Todd~D.
  Day, Justin~R. Sparks, Hiu~Y. Cheng, John~V. Badding, and Anna~C. Peacock.
\newblock {Extreme electronic bandgap modification in laser-crystallized
  silicon optical fibres}.
\newblock \emph{Nature Materials}, 13\penalty0 (12):\penalty0 1122--1127, 2014.
\newblock ISSN 14764660.
\newblock \doi{10.1038/nmat4098}.

\bibitem[Katiyar et~al.(2020)Katiyar, Thai, Yun, Lee, and Ahn]{Katiyar2020}
Ajit~K. Katiyar, Kean~You Thai, Won~Seok Yun, Jae~Dong Lee, and Jong~Hyun Ahn.
\newblock {Breaking the absorption limit of Si toward SWIR wavelength range via
  strain engineering}.
\newblock \emph{Science Advances}, 6\penalty0 (31):\penalty0 1--11, 2020.
\newblock ISSN 23752548.
\newblock \doi{10.1126/sciadv.abb0576}.

\bibitem[Dresselhaus et~al.(1955)Dresselhaus, Kip, and Kittel]{Dresselhaus1955}
G.~Dresselhaus, A.~F. Kip, and C.~Kittel.
\newblock Plasma resonance in crystals: Observations and theory.
\newblock \emph{Phys. Rev.}, 100:\penalty0 618--625, Oct 1955.
\newblock \doi{10.1103/PhysRev.100.618}.
\newblock URL \url{https://link.aps.org/doi/10.1103/PhysRev.100.618}.

\bibitem[Rideau et~al.(2006)Rideau, Feraille, Ciampolini, Minondo, Tavernier,
  Jaouen, and Ghetti]{Rideau2006}
D.~Rideau, M.~Feraille, L.~Ciampolini, M.~Minondo, C.~Tavernier, H.~Jaouen, and
  A.~Ghetti.
\newblock Strained si, ge, and
  ${\mathrm{si}}_{1\ensuremath{-}x}{\mathrm{ge}}_{x}$ alloys modeled with a
  first-principles-optimized full-zone $k \cdot p$ method.
\newblock \emph{Phys. Rev. B}, 74:\penalty0 195208, Nov 2006.
\newblock \doi{10.1103/PhysRevB.74.195208}.
\newblock URL \url{https://link.aps.org/doi/10.1103/PhysRevB.74.195208}.

\bibitem[Chelikowsky and Cohen(1976)]{Chelikowsky1976}
James~R. Chelikowsky and Marvin~L. Cohen.
\newblock {Nonlocal pseudopotential calculations for the electronic structure
  of eleven diamond and zinc-blende semiconductors}.
\newblock \emph{Physical Review B}, 14\penalty0 (2):\penalty0 556--582, 1976.
\newblock ISSN 01631829.
\newblock \doi{10.1103/PhysRevB.14.556}.

\bibitem[Fischetti and Laux(1996)]{Fischetti1996}
M.~V. Fischetti and S.~E. Laux.
\newblock {Band structure, deformation potentials, and carrier mobility in
  strained Si, Ge, and SiGe alloys}.
\newblock \emph{Journal of Applied Physics}, 80\penalty0 (4):\penalty0
  2234--2252, aug 1996.
\newblock ISSN 0021-8979.
\newblock \doi{10.1063/1.363052}.
\newblock URL \url{http://aip.scitation.org/doi/10.1063/1.363052}.

\bibitem[Slater and Koster(1954)]{Slater1954}
J.~C. Slater and G.~F. Koster.
\newblock Simplified lcao method for the periodic potential problem.
\newblock \emph{Phys. Rev.}, 94:\penalty0 1498--1524, Jun 1954.
\newblock \doi{10.1103/PhysRev.94.1498}.
\newblock URL \url{https://link.aps.org/doi/10.1103/PhysRev.94.1498}.

\bibitem[Niquet et~al.(2009)Niquet, Rideau, Tavernier, Jaouen, and
  Blase]{Niquet2009}
Y.~M. Niquet, D.~Rideau, C.~Tavernier, H.~Jaouen, and X.~Blase.
\newblock Onsite matrix elements of the tight-binding hamiltonian of a strained
  crystal: Application to silicon, germanium, and their alloys.
\newblock \emph{Phys. Rev. B}, 79:\penalty0 245201, Jun 2009.
\newblock \doi{10.1103/PhysRevB.79.245201}.
\newblock URL \url{https://link.aps.org/doi/10.1103/PhysRevB.79.245201}.

\bibitem[Bouhassoune and Schindlmayr(2015)]{Bouhassoune2015}
Mohammed Bouhassoune and Arno Schindlmayr.
\newblock {Ab initio study of strain effects on the quasiparticle bands and
  effective masses in silicon}.
\newblock \emph{Advances in Condensed Matter Physics}, 2015, 2015.
\newblock ISSN 16878124.
\newblock \doi{10.1155/2015/453125}.

\bibitem[Wen and Bellotti(2015)]{Wen2015}
Hanqing Wen and Enrico Bellotti.
\newblock {Rigorous theory of the radiative and gain characteristics of silicon
  and germanium lasing media}.
\newblock \emph{Physical Review B}, 91\penalty0 (3), jan 2015.
\newblock ISSN 1098-0121.
\newblock \doi{10.1103/PhysRevB.91.035307}.
\newblock URL \url{https://link.aps.org/doi/10.1103/PhysRevB.91.035307}.

\bibitem[Shi et~al.(2019)Shi, Tsymbalov, Dao, Suresh, Shapeev, and Li]{Shi2019}
Zhe Shi, Evgenii Tsymbalov, Ming Dao, Subra Suresh, Alexander Shapeev, and
  Ju~Li.
\newblock {Deep elastic strain engineering of bandgap through machine
  learning}.
\newblock \emph{Proceedings of the National Academy of Sciences of the United
  States of America}, 116\penalty0 (10):\penalty0 4117--4122, 2019.
\newblock ISSN 10916490.
\newblock \doi{10.1073/pnas.1818555116}.

\bibitem[Janik et~al.(2020)Janik, Scharoch, and Kudrawiec]{Janik2020}
Norbert Janik, Pawel Scharoch, and Robert Kudrawiec.
\newblock Towards band gap engineering via biaxial and axial strain in group iv
  crystals.
\newblock \emph{Computational Materials Science}, 181:\penalty0 109729, 2020.
\newblock ISSN 0927-0256.
\newblock \doi{https://doi.org/10.1016/j.commatsci.2020.109729}.
\newblock URL
  \url{https://www.sciencedirect.com/science/article/pii/S0927025620302202}.

\bibitem[Cavallo and Lagally(2012)]{Cavallo2012}
Francesca Cavallo and Max~G. Lagally.
\newblock {Semiconductor nanomembranes: A platform for new properties via
  strain engineering}.
\newblock \emph{Nanoscale Research Letters}, 7:\penalty0 1--10, 2012.
\newblock ISSN 1556276X.
\newblock \doi{10.1186/1556-276X-7-628}.

\bibitem[Reiche et~al.(2009)Reiche, Moutanabbir, Hoentschel, G{\"{o}}sele,
  Flachowsky, and Horstmann]{Reiche2009}
Manfred Reiche, O.~Moutanabbir, Jan Hoentschel, U.M. G{\"{o}}sele, Stefan
  Flachowsky, and Manfred Horstmann.
\newblock {Strained Silicon Devices}.
\newblock \emph{Solid State Phenomena}, 156-158:\penalty0 61--68, oct 2009.
\newblock ISSN 1662-9779.
\newblock \doi{10.4028/www.scientific.net/SSP.156-158.61}.
\newblock URL \url{https://www.scientific.net/SSP.156-158.61}.

\bibitem[Wang et~al.(2008)Wang, Yang, Morioka, Kitamura, and
  Nakashima]{Wang2008}
Dong Wang, Haigui Yang, Jun Morioka, Tokuhide Kitamura, and Hiroshi Nakashima.
\newblock {Local strain evaluation for freestanding Si membranes by
  microphotoluminescence using UV laser excitation}.
\newblock In \emph{2008 9th International Conference on Solid-State and
  Integrated-Circuit Technology}, pages 684--687. IEEE, oct 2008.
\newblock ISBN 978-1-4244-2185-5.
\newblock \doi{10.1109/ICSICT.2008.4734646}.
\newblock URL \url{http://ieeexplore.ieee.org/document/4734646/}.

\bibitem[Wang et~al.(2010)Wang, Yang, Kitamura, and Nakashima]{Wang2010}
Dong Wang, Haigui Yang, Tokuhide Kitamura, and Hiroshi Nakashima.
\newblock {Influence of freely diffusing excitons on the photoluminescence
  spectrum of Si thick films with depth distribution of strain}.
\newblock \emph{Journal of Applied Physics}, 107\penalty0 (3), feb 2010.
\newblock ISSN 0021-8979.
\newblock \doi{10.1063/1.3305463}.
\newblock URL \url{http://aip.scitation.org/doi/10.1063/1.3305463}.

\bibitem[Mungu{\'{i}}a et~al.(2008{\natexlab{a}})Mungu{\'{i}}a, Bluet, Baira,
  Marty, Bremond, Hartmann, and Mermoux]{Munguia2008a}
J.~Mungu{\'{i}}a, J-M. Bluet, M.~Baira, O.~Marty, G.~Bremond, J.~M. Hartmann,
  and M.~Mermoux.
\newblock {Thickness dependence of photoluminescence for tensely strained
  silicon layer on insulator}.
\newblock \emph{Applied Physics Letters}, 93\penalty0 (19), nov
  2008{\natexlab{a}}.
\newblock ISSN 0003-6951.
\newblock \doi{10.1063/1.3023058}.
\newblock URL \url{http://aip.scitation.org/doi/10.1063/1.3023058}.

\bibitem[Mungu{\'{i}}a et~al.(2008{\natexlab{b}})Mungu{\'{i}}a, Bremond, Bluet,
  Hartmann, and Mermoux]{Munguia2008b}
J.~Mungu{\'{i}}a, G.~Bremond, J.~M. Bluet, J.~M. Hartmann, and M.~Mermoux.
\newblock {Strain dependence of indirect band gap for strained silicon on
  insulator wafers}.
\newblock \emph{Applied Physics Letters}, 93\penalty0 (10):\penalty0 19--22,
  2008{\natexlab{b}}.
\newblock ISSN 00036951.
\newblock \doi{10.1063/1.2978241}.

\bibitem[Mungu{\'{i}}a et~al.(2012)Mungu{\'{i}}a, Bluet, Marty, Bremond,
  Mermoux, and Rouchon]{Munguia2012}
J.~Mungu{\'{i}}a, J.~M. Bluet, O.~Marty, G.~Bremond, M.~Mermoux, and
  D.~Rouchon.
\newblock {Temperature dependence of the indirect bandgap in ultrathin strained
  silicon on insulator layer}.
\newblock \emph{Applied Physics Letters}, 100\penalty0 (10), 2012.
\newblock ISSN 00036951.
\newblock \doi{10.1063/1.3691955}.

\bibitem[Heremans et~al.(2016)Heremans, Tripathi, {de Jamblinne de Meux},
  Smits, Hou, Pourtois, and Gelinck]{Heremans2016}
Paul Heremans, Ashutosh~K. Tripathi, Albert {de Jamblinne de Meux}, Edsger~C.P.
  Smits, Bo~Hou, Geoffrey Pourtois, and Gerwin~H. Gelinck.
\newblock {Mechanical and Electronic Properties of Thin-Film Transistors on
  Plastic, and Their Integration in Flexible Electronic Applications}.
\newblock \emph{Advanced Materials}, 28\penalty0 (22):\penalty0 4266--4282,
  2016.
\newblock ISSN 15214095.
\newblock \doi{10.1002/adma.201504360}.

\bibitem[Montmeat et~al.(2016)Montmeat, {De Nigris Brandolisi}, Tardif, Enot,
  Enyedi, Kachtouli, Besson, Rieutord, and Fournel]{Montmeat2016}
P.~Montmeat, I.~{De Nigris Brandolisi}, S.~Tardif, T.~Enot, G.~Enyedi,
  R.~Kachtouli, P.~Besson, F.~Rieutord, and F.~Fournel.
\newblock {Transfer of Ultra-Thin Semi-Conductor Films onto Flexible
  Substrates}.
\newblock \emph{ECS Transactions}, 75\penalty0 (9):\penalty0 247--252, sep
  2016.
\newblock ISSN 1938-6737.
\newblock \doi{10.1149/07509.0247ecst}.
\newblock URL \url{https://iopscience.iop.org/article/10.1149/07509.0247ecst}.

\bibitem[S{\'a}nchez-P{\'e}rez et~al.(2011)S{\'a}nchez-P{\'e}rez, Boztug, Chen,
  Sudradjat, Paskiewicz, Jacobson, Lagally, and Paiella]{sanchez2011direct}
Jose~R S{\'a}nchez-P{\'e}rez, Cicek Boztug, Feng Chen, Faisal~F Sudradjat,
  Deborah~M Paskiewicz, RB~Jacobson, Max~G Lagally, and Roberto Paiella.
\newblock Direct-bandgap light-emitting germanium in tensilely strained
  nanomembranes.
\newblock \emph{Proceedings of the National Academy of Sciences}, 108\penalty0
  (47):\penalty0 18893--18898, 2011.

\bibitem[Gassenq et~al.(2015)Gassenq, Guilloy, Osvaldo~Dias, Pauc, Rouchon,
  Hartmann, Widiez, Tardif, Rieutord, Escalante, et~al.]{gassenq20151}
Alban Gassenq, Kevin Guilloy, Guilherme Osvaldo~Dias, Nicolas Pauc, Denis
  Rouchon, J-M Hartmann, Julie Widiez, Samuel Tardif, Fran{\c{c}}ois Rieutord,
  Jos{\'e} Escalante, et~al.
\newblock 1.9\% bi-axial tensile strain in thick germanium suspended membranes
  fabricated in optical germanium-on-insulator substrates for laser
  applications.
\newblock \emph{Applied Physics Letters}, 107\penalty0 (19):\penalty0 191904,
  2015.

\bibitem[Passi et~al.(2012)Passi, Bhaskar, Pardoen, Sodervall, Nilsson,
  Petersson, Hagberg, and Raskin]{Passi2012}
Vikram Passi, Umesh Bhaskar, Thomas Pardoen, Ulf Sodervall, Bengt Nilsson,
  G{\"{o}}ran Petersson, Mats Hagberg, and Jean-Pierre Raskin.
\newblock {High-Throughput On-Chip Large Deformation of Silicon Nanoribbons and
  Nanowires}.
\newblock \emph{Journal of Microelectromechanical Systems}, 21\penalty0
  (4):\penalty0 822--829, aug 2012.
\newblock ISSN 1057-7157.
\newblock \doi{10.1109/JMEMS.2012.2190711}.
\newblock URL \url{http://ieeexplore.ieee.org/document/6179958/}.

\bibitem[{Kumar Bhaskar} et~al.(2013){Kumar Bhaskar}, Pardoen, Passi, and
  Raskin]{KumarBhaskar2013}
Umesh {Kumar Bhaskar}, Thomas Pardoen, Vikram Passi, and Jean-Pierre Raskin.
\newblock {Piezoresistance of nano-scale silicon up to 2 GPa in tension}.
\newblock \emph{Applied Physics Letters}, 102\penalty0 (3):\penalty0 031911,
  jan 2013.
\newblock ISSN 0003-6951.
\newblock \doi{10.1063/1.4788919}.
\newblock URL \url{http://aip.scitation.org/doi/10.1063/1.4788919}.

\bibitem[Ure{\~{n}}a et~al.(2012)Ure{\~{n}}a, Olsen, {\v{S}}iller, Bhaskar,
  Pardoen, and Raskin]{Urena2012}
Ferran Ure{\~{n}}a, Sarah~H. Olsen, Lidija {\v{S}}iller, Umesh Bhaskar, Thomas
  Pardoen, and Jean-Pierre Raskin.
\newblock {Strain in silicon nanowire beams}.
\newblock \emph{Journal of Applied Physics}, 112\penalty0 (11):\penalty0
  114506, dec 2012.
\newblock ISSN 0021-8979.
\newblock \doi{10.1063/1.4765025}.
\newblock URL \url{http://aip.scitation.org/doi/10.1063/1.4765025}.

\bibitem[Zhang et~al.(2016)Zhang, Tersoff, Xu, Chen, Zhang, Zhang, Yang, Lee,
  Tu, Li, and Lu]{Zhang2016}
Hongti Zhang, Jerry Tersoff, Shang Xu, Huixin Chen, Qiaobao Zhang, Kaili Zhang,
  Yong Yang, Chun-Sing Lee, King-Ning Tu, Ju~Li, and Yang Lu.
\newblock {Approaching the ideal elastic strain limit in silicon nanowires}.
\newblock \emph{Science Advances}, 2\penalty0 (8), aug 2016.
\newblock ISSN 2375-2548.
\newblock \doi{10.1126/sciadv.1501382}.
\newblock URL
  \url{https://advances.sciencemag.org/lookup/doi/10.1126/sciadv.1501382
  https://www.science.org/doi/10.1126/sciadv.1501382}.

\bibitem[Gravier et~al.(2009)Gravier, Coulombier, Safi, Andre, Boe, Raskin, and
  Pardoen]{Gravier2009}
S.~Gravier, M.~Coulombier, A.~Safi, N.~Andre, A.~Boe, J.-P. Raskin, and
  T.~Pardoen.
\newblock {New On-Chip Nanomechanical Testing Laboratory - Applications to
  Aluminum and Polysilicon Thin Films}.
\newblock \emph{Journal of Microelectromechanical Systems}, 18\penalty0
  (3):\penalty0 555--569, jun 2009.
\newblock ISSN 1057-7157.
\newblock \doi{10.1109/JMEMS.2009.2020380}.
\newblock URL \url{http://ieeexplore.ieee.org/document/4957038/}.

\bibitem[Pardoen et~al.(2016)Pardoen, Colla, Idrissi, Amin-Ahmadi, Wang,
  Schryvers, Bhaskar, and Raskin]{PARDOEN2016485}
Thomas Pardoen, Marie-Stéphane Colla, Hosni Idrissi, Behnam Amin-Ahmadi,
  Binjie Wang, Dominique Schryvers, Umesh~K. Bhaskar, and Jean-Pierre Raskin.
\newblock A versatile lab-on-chip test platform to characterize elementary
  deformation mechanisms and electromechanical couplings in nanoscopic objects.
\newblock \emph{Comptes Rendus Physique}, 17\penalty0 (3):\penalty0 485--495,
  2016.
\newblock ISSN 1631-0705.
\newblock \doi{https://doi.org/10.1016/j.crhy.2015.11.005}.
\newblock URL
  \url{https://www.sciencedirect.com/science/article/pii/S1631070515002224}.
\newblock Physique de la matière condensée au XXIe siècle: l’héritage de
  Jacques Friedel.

\bibitem[Bhaskar et~al.(2012)Bhaskar, Passi, Houri, Escobedo-Cousin, Olsen,
  Pardoen, and Raskin]{Bhaskar2011}
Umesh Bhaskar, Vikram Passi, Samer Houri, Enrique Escobedo-Cousin, Sarah~H
  Olsen, Thomas Pardoen, and Jean-Pierre Raskin.
\newblock {On-chip tensile testing of nanoscale silicon free-standing beams}.
\newblock \emph{Journal of Materials Research}, 27\penalty0 (3):\penalty0
  571--579, feb 2012.
\newblock ISSN 0884-2914.
\newblock \doi{10.1557/jmr.2011.340}.

\bibitem[Ganesan et~al.(1970)Ganesan, Maradudin, and Oitmaa]{Ganesan1970}
S~Ganesan, A~A Maradudin, and J~Oitmaa.
\newblock {A lattice theory of morphic effects in crystals of the diamond
  structure}.
\newblock \emph{Annals of Physics}, 56\penalty0 (2):\penalty0 556--594, 1970.
\newblock ISSN 0003-4916.
\newblock \doi{https://doi.org/10.1016/0003-4916(70)90029-1}.
\newblock URL
  \url{https://www.sciencedirect.com/science/article/pii/0003491670900291}.

\bibitem[Roisin et~al.(2023)Roisin, Colla, Raskin, and Flandre]{Roisin2023}
Nicolas Roisin, Marie~St{\'{e}}phane Colla, Jean~Pierre Raskin, and Denis
  Flandre.
\newblock {Raman strain–shift measurements and prediction from
  first-principles in highly strained silicon}.
\newblock \emph{Journal of Materials Science: Materials in Electronics},
  34\penalty0 (5), 2023.
\newblock ISSN 1573482X.
\newblock \doi{10.1007/s10854-022-09769-3}.

\bibitem[Nguyen and Macdonald(2017)]{Nguyen2017}
Hieu~T. Nguyen and Daniel Macdonald.
\newblock {On the composition of luminescence spectra from heavily doped p-type
  silicon under low and high excitation}.
\newblock \emph{Journal of Luminescence}, 181:\penalty0 223--229, 2017.
\newblock ISSN 00222313.
\newblock \doi{10.1016/j.jlumin.2016.08.036}.
\newblock URL \url{http://dx.doi.org/10.1016/j.jlumin.2016.08.036}.

\bibitem[Wagner(1984)]{Wagner1984}
J.~Wagner.
\newblock {Photoluminescence and excitation spectroscopy in heavily doped n-
  and p-type silicon}.
\newblock \emph{Physical Review B}, 29\penalty0 (4):\penalty0 2002--2009, 1984.
\newblock ISSN 01631829.
\newblock \doi{10.1103/PhysRevB.29.2002}.

\bibitem[Yoo et~al.(2015)Yoo, Kang, Murai, and Yoshimoto]{Yoo2015}
Woo~Sik Yoo, Kitaek Kang, Gota Murai, and Masahiro Yoshimoto.
\newblock {Temperature Dependence of Photoluminescence Spectra from Crystalline
  Silicon}.
\newblock \emph{ECS Journal of Solid State Science and Technology}, 4\penalty0
  (12):\penalty0 456--461, 2015.
\newblock ISSN 2162-8769.
\newblock \doi{10.1149/2.0251512jss}.

\bibitem[Okada and Tokumaru(1984)]{Okada1984}
Yasumasa Okada and Yozo Tokumaru.
\newblock {Precise determination of lattice parameter and thermal expansion
  coefficient of silicon between 300 and 1500 K}.
\newblock \emph{Journal of Applied Physics}, 56\penalty0 (2):\penalty0
  314--320, 1984.
\newblock ISSN 00218979.
\newblock \doi{10.1063/1.333965}.

\bibitem[Madelung(1991)]{Madelung1991}
O.~Madelung.
\newblock \emph{{Semiconductors}}.
\newblock Data in Science and Technology. Springer Berlin Heidelberg, Berlin,
  Heidelberg, 1991.
\newblock ISBN 978-3-540-53150-0.
\newblock \doi{10.1007/978-3-642-45681-7}.
\newblock URL \url{http://link.springer.com/10.1007/978-3-642-45681-7
  https://link.springer.com/10.1007/978-3-642-45681-7}.

\bibitem[Gonze et~al.(2020)Gonze, Amadon, Antonius, Arnardi, Baguet, Beuken,
  Bieder, Bottin, Bouchet, Bousquet, et~al.]{Gonze2020}
Xavier Gonze, Bernard Amadon, Gabriel Antonius, Fr{\'e}d{\'e}ric Arnardi, Lucas
  Baguet, Jean-Michel Beuken, Jordan Bieder, Fran{\c{c}}ois Bottin, Johann
  Bouchet, Eric Bousquet, et~al.
\newblock The abinit project: Impact, environment and recent developments.
\newblock \emph{Computer Physics Communications}, 248:\penalty0 107042, 2020.
\newblock \doi{https://doi.org/10.1016/j.cpc.2019.107042}.

\bibitem[Romero et~al.(2020)Romero, Allan, Amadon, Antonius, Applencourt,
  Baguet, Bieder, Bottin, Bouchet, Bousquet, et~al.]{Romero2020}
Aldo~H Romero, Douglas~C Allan, Bernard Amadon, Gabriel Antonius, Thomas
  Applencourt, Lucas Baguet, Jordan Bieder, Fran{\c{c}}ois Bottin, Johann
  Bouchet, Eric Bousquet, et~al.
\newblock Abinit: Overview and focus on selected capabilities.
\newblock \emph{The Journal of chemical physics}, 152\penalty0 (12):\penalty0
  124102, 2020.
\newblock \doi{https://doi.org/10.1063/1.5144261}.

\bibitem[Van~Setten et~al.(2018)Van~Setten, Giantomassi, Bousquet, Verstraete,
  Hamann, Gonze, and Rignanese]{Vansetten2018}
MJ~Van~Setten, Matteo Giantomassi, Eric Bousquet, Matthieu~J Verstraete, Don~R
  Hamann, Xavier Gonze, and G-M Rignanese.
\newblock The pseudodojo: Training and grading a 85 element optimized
  norm-conserving pseudopotential table.
\newblock \emph{Computer Physics Communications}, 226:\penalty0 39--54, 2018.
\newblock \doi{https://doi.org/10.1016/j.cpc.2018.01.012}.

\bibitem[Perdew et~al.(1996)Perdew, Burke, and Ernzerhof]{Perdew1996}
John~P. Perdew, Kieron Burke, and Matthias Ernzerhof.
\newblock {Generalized Gradient Approximation Made Simple}.
\newblock \emph{Physical Review Letters}, 77\penalty0 (18):\penalty0
  3865--3868, oct 1996.
\newblock ISSN 0031-9007.
\newblock \doi{10.1103/PhysRevLett.77.3865}.
\newblock URL \url{https://link.aps.org/doi/10.1103/PhysRevLett.77.3865}.

\bibitem[Monkhorst and Pack(1976)]{Monkhorst1976}
Hendrik~J Monkhorst and James~D Pack.
\newblock Special points for brillouin-zone integrations.
\newblock \emph{Physical Review B}, 13\penalty0 (12):\penalty0 5188, 1976.
\newblock \doi{https://doi.org/10.1103/PhysRevB.13.5188}.

\bibitem[Onida et~al.(2002)Onida, Reining, and Rubio]{Onida2002}
Giovanni Onida, Lucia Reining, and Angel Rubio.
\newblock {Electronic excitations: density-functional versus many-body
  Green's-function approaches}.
\newblock \emph{Reviews of Modern Physics}, 74\penalty0 (2):\penalty0 601--659,
  jun 2002.
\newblock ISSN 0034-6861.
\newblock \doi{10.1103/RevModPhys.74.601}.
\newblock URL
  \url{https://linkinghub.elsevier.com/retrieve/pii/S0927024804003502
  https://link.aps.org/doi/10.1103/RevModPhys.74.601}.

\bibitem[Faleev et~al.(2004)Faleev, {Van Schilfgaarde}, and Kotani]{Faleev2004}
Sergey~V. Faleev, Mark {Van Schilfgaarde}, and Takao Kotani.
\newblock {All-electron self-consistent GW approximation: Application to Si,
  MnO, and NiO}.
\newblock \emph{Physical Review Letters}, 93\penalty0 (12):\penalty0 12--15,
  2004.
\newblock ISSN 00319007.
\newblock \doi{10.1103/PhysRevLett.93.126406}.

\bibitem[Escobedo-Cousin et~al.(2011)Escobedo-Cousin, Olsen, Pardoen, Bhaskar,
  and Raskin]{Escobedo-Cousin2011}
E.~Escobedo-Cousin, S.~H. Olsen, T.~Pardoen, U.~Bhaskar, and J.-P. Raskin.
\newblock {Experimental observations of surface roughness in uniaxially loaded
  strained Si microelectromechanical systems-based structures}.
\newblock \emph{Applied Physics Letters}, 99\penalty0 (24):\penalty0 241906,
  dec 2011.
\newblock ISSN 0003-6951.
\newblock \doi{10.1063/1.3669413}.
\newblock URL \url{http://aip.scitation.org/doi/10.1063/1.3669413}.

\bibitem[Colla et~al.(2015)Colla, Amin-Ahmadi, Idrissi, Malet, Godet, Raskin,
  Schryvers, and Pardoen]{Colla2015}
M.~S. Colla, B.~Amin-Ahmadi, H.~Idrissi, L.~Malet, S.~Godet, J.~P. Raskin,
  D.~Schryvers, and T.~Pardoen.
\newblock Dislocation-mediated relaxation in nanograined columnar palladium
  films revealed by on-chip time-resolved hrtem testing.
\newblock \emph{Nature Communications}, 6\penalty0 (1):\penalty0 5922, January
  2015.
\newblock ISSN 2041-1723.
\newblock URL \url{https://doi.org/10.1038/ncomms6922}.

\bibitem[{De Wolf}(1996)]{DeWolf1996}
Ingrid {De Wolf}.
\newblock {Micro-Raman spectroscopy to study local mechanical stress in silicon
  integrated circuits}.
\newblock \emph{Semiconductor Science and Technology}, 11\penalty0
  (2):\penalty0 139--154, 1996.
\newblock ISSN 02681242.
\newblock \doi{10.1088/0268-1242/11/2/001}.

\bibitem[Loudon(1964)]{Loudon1964}
R~Loudon.
\newblock {The Raman effect in crystals}.
\newblock \emph{Advances in Physics}, 13\penalty0 (52):\penalty0 423--482, oct
  1964.
\newblock ISSN 0001-8732.
\newblock \doi{10.1080/00018736400101051}.
\newblock URL \url{https://doi.org/10.1080/00018736400101051}.

\bibitem[Pelik{\'{a}}n et~al.(2020)Pelik{\'{a}}n, {\v{C}}eppan, and
  Li{\v{s}}ka]{Pelikan1994}
Peter Pelik{\'{a}}n, Michal {\v{C}}eppan, and Marek Li{\v{s}}ka.
\newblock \emph{{Applications of Numerical Methods in Molecular Spectroscopy}}.
\newblock CRC Press, nov 2020.
\newblock ISBN 9781003068686.
\newblock \doi{10.1201/9781003068686}.
\newblock URL \url{https://www.taylorfrancis.com/books/9781000098945}.

\bibitem[Ure{\~{n}}a et~al.(2013)Ure{\~{n}}a, Olsen, and Raskin]{Urena2013}
Ferran Ure{\~{n}}a, Sarah~H. Olsen, and Jean~Pierre Raskin.
\newblock {Raman measurements of uniaxial strain in silicon nanostructures}.
\newblock \emph{Journal of Applied Physics}, 114\penalty0 (14), 2013.
\newblock ISSN 00218979.
\newblock \doi{10.1063/1.4824291}.

\bibitem[Rau and Werner(2004)]{Rau2004}
U.~Rau and J.~H. Werner.
\newblock {Radiative efficiency limits of solar cells with lateral band-gap
  fluctuations}.
\newblock \emph{Applied Physics Letters}, 84\penalty0 (19):\penalty0
  3735--3737, 04 2004.
\newblock ISSN 0003-6951.
\newblock \doi{10.1063/1.1737071}.
\newblock URL \url{https://doi.org/10.1063/1.1737071}.

\bibitem[Mattheis et~al.(2007)Mattheis, Rau, and Werner]{Mattheis2007}
Julian Mattheis, Uwe Rau, and J{\"{u}}rgen~H. Werner.
\newblock {Light absorption and emission in semiconductors with band gap
  fluctuations—A study on Cu(In,Ga)Se2 thin films}.
\newblock \emph{Journal of Applied Physics}, 101, jun 2007.
\newblock ISSN 0021-8979.
\newblock \doi{10.1063/1.2721768}.
\newblock URL \url{http://aip.scitation.org/doi/10.1063/1.2721768}.

\end{thebibliography}


\end{document}